# Spiralling molecular structures and chiral selectivity in model membranes

Alexander P. Fellows[1], Ben John[1], Martin Wolf[1], and Martin Thämer[1]*

[1]Fritz-Haber-Institut der Max-Planck-Gesellschaft, Faradayweg 4-6, 14195, Berlin, Germany

*Corresponding Author

thaemer@fhi-berlin.mpg.de

## Summary

Since the lipid raft model was developed at the end of the last century, it became clear that the specific molecular arrangements of phospholipid assemblies within a membrane have profound implications in a vast range of physiological functions.[1–4] Studies of such condensed lipid islands in model systems using fluorescence and Brewster angle microscopies have shown a wide range of sizes and morphologies, with suggestions of substantial in-plane molecular anisotropy and mesoscopic structural chirality.[5–8] Whilst these variations can significantly alter many membrane properties including its fluidity, permeability, and molecular recognition, the details of the in-plane molecular orientations underlying these traits remain largely unknown.[3,9] Here, we use phase-resolved sum-frequency generation microscopy on model membranes of phospholipid monolayers with mixed molecular chirality, which form micron-scale circular domains of condensed lipids, to fully determine their three-dimensional molecular structure. We find that the domains possess curved molecular directionality with spiralling mesoscopic packing. By comparing different enantiomeric mixtures, both the molecular and spiral turning directions are shown to depend on the lipid chirality, but with a clear deviation from mirror symmetry in the formed structures. This demonstrates strong enantioselectivity in the domain growth process, which has potential connections to the evolution of homochirality in all living organisms[10] as well as implications for enantioselective drug design[11].

# Introduction

The importance of phospholipids in biology cannot be overstated. Their amphiphilic nature promotes self-assembly into two-dimensional membranes where the hydrophobic tails tightly pack and point away from the membrane interface (see Fig. 1a).[12] This well-defined molecular orientation normal to the membrane leaflet is the basis of their fundamental physiological purpose as barriers separating different environments within an organism. Biological membranes contain, among many other constituents, a complex mixture of saturated and unsaturated chiral lipids and form highly specialised regions that are involved in a plethora of physiological processes such as endo- and exocytosis, cell signalling, and cellular adhesion or recognition.[3,13] An example of such a specialized region that has recently attracted much attention are lipid rafts, which are co-existing phases of condensed lipid islands within a liquid-like phase. Lipid rafts have been linked to many of the aforementioned membrane functions but their exact nature and molecular structure is largely unknown, and their role in biological processes highly debated.[11,14–17]

One key to a better understanding of the formation, and ultimately the function of lipid rafts is encoded in their in-plane molecular structure that largely governs their physico-chemical properties. The degree of in-plane order is connected to the lipid phase behaviour, with liquid-expanded (LE) phases generally show little, while liquid-condensed (LC) regions, such as lipid rafts, possess high lateral ordering.[18] Equally important for the properties of these condensed domains are the details of the underlying molecular structure. Owing to the asymmetric in-plane shape of phospholipid molecules (Fig. 1b) there are various possibilities how they could be laterally assembled and arranged (examples are shown in Fig. 1b). Which arrangement exist under which condition is largely unknown due to the lack of microscopic insight. Furthermore, it is unclear to what extend lipid chirality influences the molecular arrangement within these condensed domains.[19,20] Enantioselectivity in the formation of lipid assemblies, resulting in different structures in homo- and heterochiral membranes, has been suggested amongst numerous tentative explanations for how and why evolution drove towards membrane homochirality in all branches of life.[10,21,22] Conclusive experimental evidence for such enantiomeric effects remains elusive. Clearly, a full molecular level picture of physiological membranes would be desirable, however, such elucidation has proven difficult.[11] Nevertheless, conceptual studies of these lipid ordering effects in model systems already represent important steps towards understanding the role lipid rafts play in biological systems.

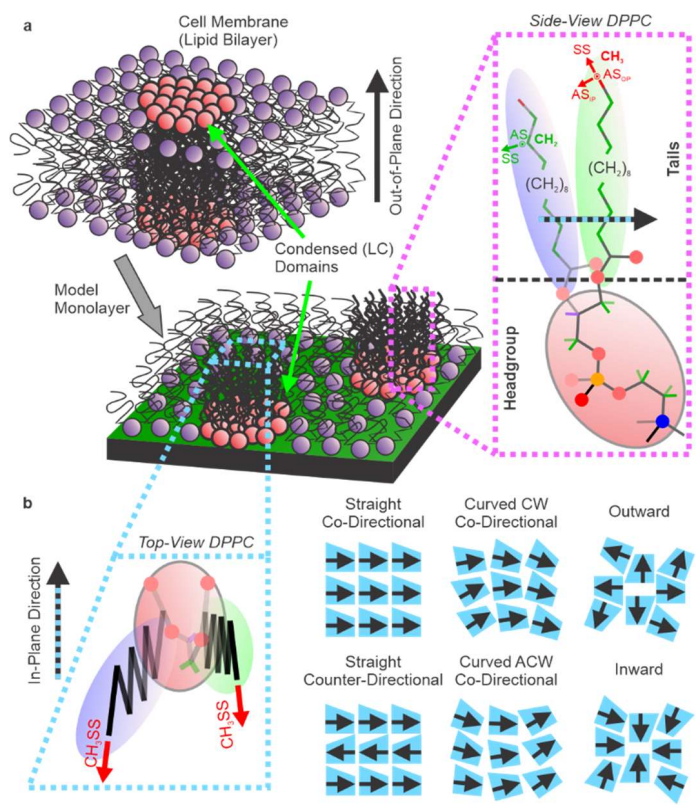

*Fig 1 Schematic structure of a phospholipid and different possible lateral packing structures. **a**, A schematic of a lipid bilayer membrane with liquid-condensed (LC) domains showing their well-defined out-of-plane directionality (black arrow). Also presented is a schematic of the LC domains formed in a model monolayer as well as the side-view structure of the R-DPPC lipid. This side-view representation also shows the C-H stretch transition dipole directions and defined in-plane molecular direction (black and blue arrow). **b**, Schematic top-view structure of the R-DPPC lipid molecule presented with a trapezoidal outline as a representation of its asymmetric in-plane morphology. This trapezium, along with the defined in-plane direction (again shown as a black and blue arrow), is then used to show examples of possible in-plane packing arrangements. CW: clockwise; ACW: anti-clockwise. Co- and counter-directional packing is defined by the specific adjacent lipid alignment directions.*

One of the most widely studied model systems is monolayers containing DPPC[a], the most abundant phospholipid in cell membranes.[23–26] Upon compression using a Langmuir trough, monolayers of both pure DPPC and mixtures with other lipids (primarily unsaturated) form LC-LE coexistence equilibria with large DPPC-rich LC domains, akin to lipid rafts. These domains have been studied using various imaging techniques including fluorescence, Brewster angle, and atomic force microscopies.[5,6,27,28] This has yielded a substantial insight into their growth, stability, and morphology. For pure DPPC monolayers, for example, the formation of domains with spiralling morphologies has been observed, with clear mirror symmetry between the two enantiomers.[7,8,29] Such morphologies suggest a complex structural hierarchy possessing a well-ordered, crystalline molecular packing with a link between molecular and mesoscopic structural chirality. This interpretation has been supported by other experiments e.g. X-ray diffraction measurements showing a high degree of in-plane order.[30] Additionally, polarised fluorescence imaging[24,31] has indicated a generally curved in-plane anisotropy, however the exact distribution of molecular orientations is still unknown. Furthermore, when mixed with unsaturated chiral lipids (a more representative scenario for biological systems), the DPPC-rich domains are typically round[32,33] which raises the question whether these domains also possess a curved molecular packing and whether this is a general feature of DPPC-rich domains. An additional important question concerns the influence of the other lipids on the growth and ultimately the internal packing within the domains, and whether there is any breaking of the mirror symmetry in such mixed systems with different enantiomeric combinations of the two lipids.

---

[a] DPPC: Dipalmitoylphosphatidylcholine

To address such questions, we developed a phase-sensitive vibrational sum-frequency generation (vSFG) microscope which can fully resolve the microscopic molecular structures of such lipid monolayers. In these experiments, we probe the C-H stretching vibrations of different DPPC-rich domains in mixed monolayers of R/S-DPPC[b,c] and fully deuterated R-POPC[d] (unsaturated). Uniquely, this imaging technique combines spectroscopic selectivity to differentiate molecular species with sensitivity for molecular orientation encoded in the phase of the signal, providing full molecular detail for these studies.[34–39] Whilst such measurements have previously not been feasible due to technical challenges such as insufficient signal-to-noise ratios, our newly developed vSFG microscope employs an advanced imaging setup[40] that overcomes this limitation.

## Results

In our vSFG experiments, the input and output beams are all set to P-polarisation (with their electric fields parallel to the X'Z'-plane) implying excitation of molecular vibrations both parallel and perpendicular to the sample surface (see Fig. 2a, for further details see Methods). For lipid molecules, the generated vSFG signals in the C-H stretch region are known to almost exclusively originate from their alkyl tails.[41] As depicted in Fig. 1a, these tails are expected to be in a largely upright configuration with their terminal $CH_3$ groups all pointing upwards. For this structure, the vSFG responses should be dominated by the $CH_3$ groups and yield a positive contribution from the symmetric stretch (SS) and negative band from both antisymmetric (AS) modes, as indicated by the Z-components of their transition dipole vectors (also shown in Fig. 1a).

The obtained vSFG images are presented in Fig. 2b, stepping through the vibrational spectrum at selected frequencies, with the spatially averaged spectrum shown in Fig. 2c (black, dashed). Only the absorptive line-shapes (imaginary parts of the phase-resolved responses) are presented, with Lorentzian-shaped peaks and dips centred at the frequencies of the vibrational transitions.[42] The first image, at 2815 cm$^{-1}$, does not correspond to any vibrational resonance from DPPC and thus no contrast can be seen. Thereafter, each image corresponds to one of the DPPC C-H resonances where the roughly circular domains become apparent. These domains have diameters of ca. 10 μm and correspond to the DPPC-rich LC phase. Specifically, the images at 2875, 2935, and 2965 cm$^{-1}$ are associated with the $CH_3$ SS, its Fermi resonance (FR), and AS, respectively, showing strong positive (SS and FR) and negative (AS) responses, as expected.[42] These resonances also clearly feature in the averaged spectrum (with dashed lines indicating the frequencies associated with each image). Fig. 2c also shows spectra averaged over only the observed domain regions (red), and only in the surrounding phase (blue). From this, it is clear that the majority of the vSFG signal arises from the DPPC-rich domains, as expected, but that the LE phase also contributes a significant signal. This reveals that the phase-separation of the two lipids is obviously incomplete, i.e., that both phases contain DPPC, only in differing proportions.

The remaining two images, at 2845 and 2905 cm$^{-1}$, correspond to the $CH_2$ SS and its FR. For the well-packed upright lipid geometry, the $CH_2$ groups lie almost completely parallel to the surface and thus should mostly yield in-plane signals that are small due to the opposing directions of the $CH_2$ groups within each molecule.[42] Clearly, the $CH_2$ signals are much weaker cf. CH$_3$, following expectations. Nevertheless, the simple presence of these CH$_2$ responses inside the domains shows that the signals from the individual lipids are not cancelling within each pixel, indicating substantial co-directional molecular in-plane ordering at the sub-micron scale. Interestingly, at the CH$_2$ frequencies, many of the domains show a split positive/negative character. These amplitude variations across the domains suggests a slowly varying molecular orientation and thus a co-directional curved molecular packing (see Fig. 1b).

---

[b] R-DPPC: 1,2-dipalmitoyl-sn-glycero-3-phosphocholine
[c] S-DPPC: 2,3-dipalmitoyl-sn-glycero-1-phosphocholine
[d] R-POPC: 1-palmitoyl-2-oleoyl-glycero-3-phosphocholine

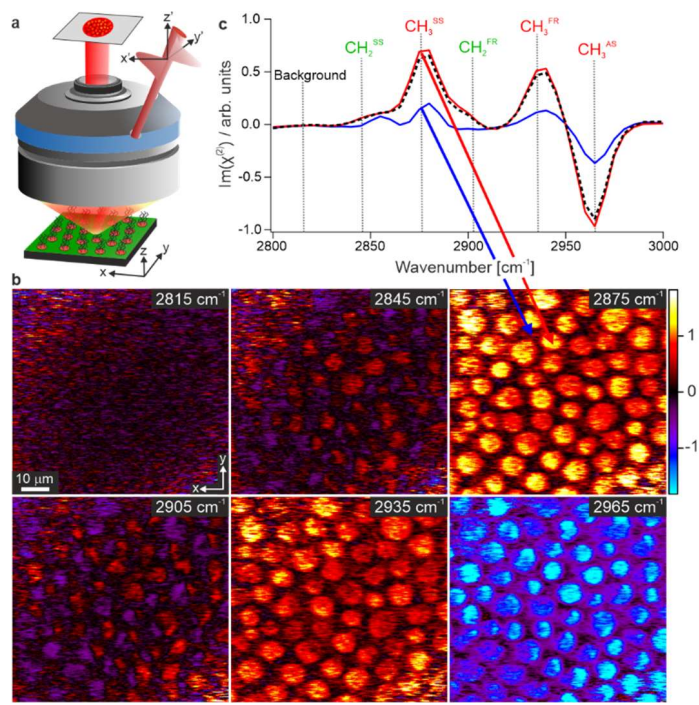

*Fig 2 vSFG microscopy images. **a**, Schematic of the phase-resolved vSFG microscope showing the input beams, lipid monolayer sample and the objective. Two coordinate systems are defined, the first (x,y,z) being the sample frame, and the second (x',y',z') being the laboratory frame that is defined based on the plane of incidence of the input beams. **b**, vSFG microscopy images of the R-DPPC/R-POPC monolayer at select frequencies. The frequencies are given in the top-right of each image and correspond to the dashed lines shown in the spectra. **c**, vSFG spectra averaged over all domains (red), outside the domains (blue), and the entire image (black, dashed). Only the imaginary (absorptive) parts of the phase-resolved responses are shown. The main CH$_2$ and CH$_3$ vibrational resonances are also indicated – SS: symmetric stretch, AS: antisymmetric stretch, FR: Fermi resonance.*

To unambiguously determine the molecular directionality, vSFG images are recorded as a function of sample rotation, yielding a four-dimensional dataset (vSFG images as a function of spectral frequency and azimuthal angle). If we consider the projection of the responses onto the probed X' and Z' directions in the laboratory frame, rotation of the sample leaves the out-of-plane responses unchanged, whereas the in-plane responses are modulated sinusoidally (see Fig. 3a). Therefore, by performing a Fourier transform (FT) of the rotations, thus converting them into azimuthal frequencies, the out-of-plane (Z') and in-plane (X') components can be separated as they appear in the 0-fold and 1-fold frequencies, respectively. A schematic of this analysis procedure is shown in Fig. 3a (further details in Supplementary Information). The resulting spatially averaged out-of-plane and in-plane spectra are given in Fig. 3b along with their deconvolution (see Methods for details) into resonances from CH$_2$ (green) and CH$_3$ (red), as well as two residual unassigned bands (blue) which likely stem from either CH$_2$ groups from the alkyl tails or the head-group, or from the single CH (methine) group present in the head-group.[25,42,43]

With the expected lipid orientation, the in-plane response should have signals from both CH$_2$ and CH$_3$ whereas the out-of-plane signal should be dominated by CH$_3$. Inspection of the deconvoluted spectra clearly shows this to be the case. Using these decompositions, we can also obtain both in-plane and out-of-plane magnitude images divided into $CH_3$ and $CH_2$ contributions (see Fig. 3c). These images confirm that the $CH_2$ groups yield strong in-plane and only minor out-of-plane signals, whereas the $CH_3$ groups give contributions to both, but are mostly present in the out-of-plane component.

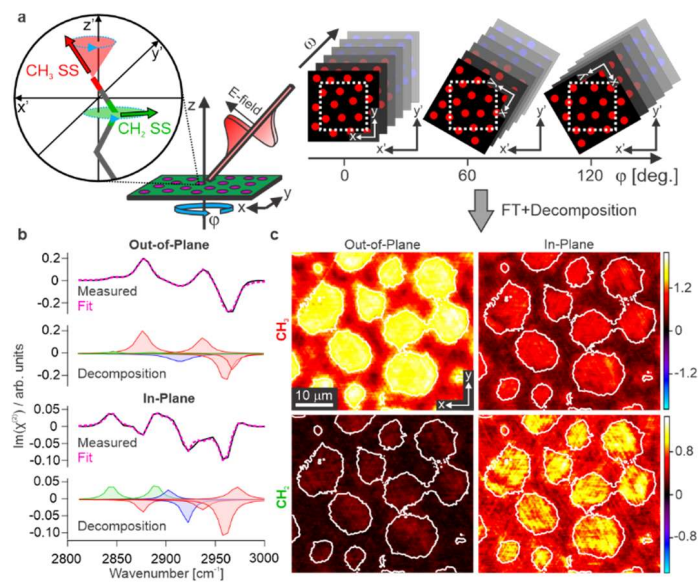

*Fig 3 Rotational analysis of vSFG microscopy images. **a**, Schematic of the rotational analysis procedure showing the variation in $CH_2$ (green) and $CH_3$ (red) transition dipoles upon sample rotation. Also shown are representations of the four-dimensional dataset that is generated, with SFG microscopy images at different spectral frequencies (ω) recorded at multiple azimuthal sample rotations (φ). The change in sample frame coordinates (x,y,z) relative to the laboratory coordinates (x',y',z') upon rotation is explicitly shown, demonstrating the need for back-rotation of the obtained images. **b**, Imaginary parts of the out-of-plane and in-plane spectra, extracted from the magnitudes of the 0-fold and 1-fold azimuthal frequencies, respectively, and averaged over all domain regions. Also shown are the deconvolutions of these spectra into their constituent bands, with $CH_2$ resonances in green, $CH_3$ resonances in red, and unassigned bands in blue. **c**, Magnitude images (in the sample frame) of the $CH_2$ and $CH_3$ resonances at the 0-fold (out-of-plane) and 1-fold (in-plane) azimuthal frequencies. The images are obtained by integrating across the frequency regions of the $CH_2$ and $CH_3$ bands. Circular contours are included in all images to highlight the domain locations. The depicted surface region is part of a wider image common to all rotations, as shown in Fig. 4a.*

With this data in hand, we can now fully determine the spatial distribution of both the out-of-plane and in-plane molecular orientations. As expected, the out-of-plane $CH_3$ signals (Fig. 3c) show a relatively homogeneous contribution for each domain which corresponds to all alkyl chains pointing up with a similar out-of-plane structure. Meanwhile, the in-plane molecular orientations are encoded in the rotational phases of the in-plane dataset, which are shown on the left side of Fig. 4b. The molecular axis that corresponds to these phases is defined according to the arrows shown in Fig. 1a and again in the colour wheel in Fig. 4b (see Supplementary Information for details).

All the details about the distribution of molecular orientations are contained within the phase image, however, translating it into a clear structural picture requires additional analysis. In a first step, the molecular directions are extracted for selected regions and displayed as in-plane arrows, as shown in Fig. 4c (left side) for an exemplary domain. From this representation, it becomes immediately clear that the lipid packing is, indeed, curved. Since the absolute molecular orientations are obtained in these measurements, the curvature direction can be unambiguously determined as clockwise (CW) for the defined molecular axis.

To evaluate if there are any higher order hierarchal structural symmetries in these domains the phase contour lines connecting locations of equal molecular orientation can be analysed. The left side of Fig. 4d shows a schematic of four possible CW curved arrangements, including both molecular direction arrows and the corresponding phase contours. Each structure creates characteristic contour line patterns (parallel, radial, or spiral with either positive or negative curvature directions) which can be used for their identification. Comparison to the contour lines extracted for the exemplary domain (also shown in Fig. 4c, left side) suggests that the mesoscopic structure of the domains is an outer subsection of a CW (+)-spiral. The formed domains thus possess a superposition of two types of curvature, namely the turning direction of molecules about the spiral centre, and the curvature of the spiral itself.

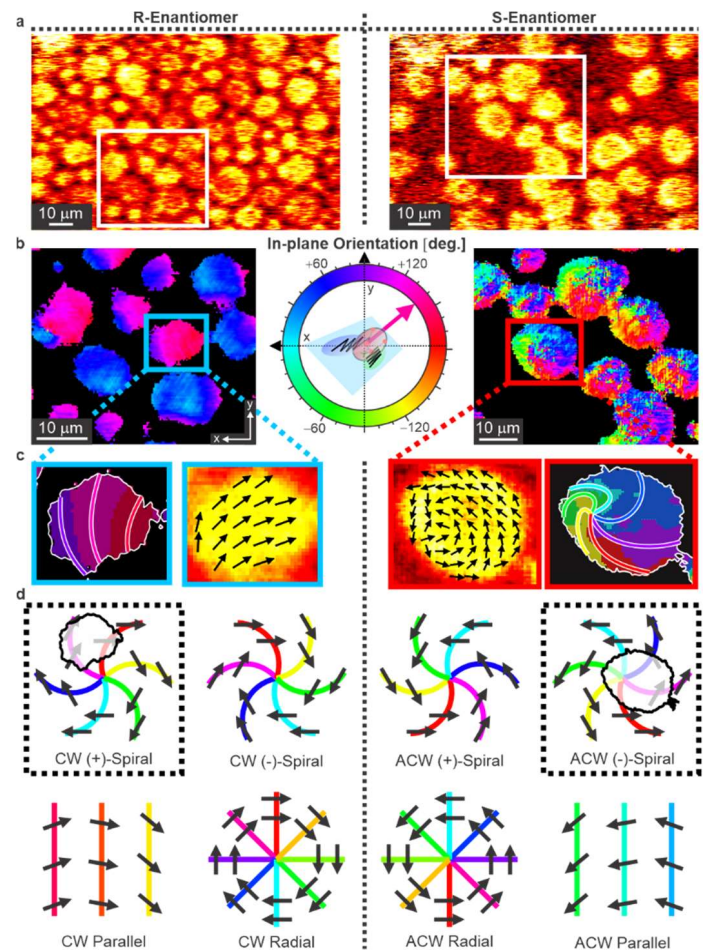

*Fig 4 In-plane phase directions of lipid rafts for each DPPC enantiomer. **a**, Overview vSFG images showing the spectrally integrated CH$_3$ magnitudes for the two monolayer samples (left: R-DPPC/R-POPC, and right: S-DPPC/R-POPC) at 0 deg. rotation angle. Also indicated are the surface regions common to all rotations (white box). **b**, Rotational phase images from the in-plane component (1-fold rotational frequency) presented as colour-maps (sample frame). The molecular directions are indicated in the phase colour wheel. **c**, Magnified image of a selected domain from each phase map highlighting the distribution of molecular directions in the form of contour lines and phase arrows. Phase arrows are obtained by averaging 6×6 pixel regions across each domain. **d**, Schematic representations of different molecular packing structures, namely spirals (where + and – indicate the spiralling direction, defined as CW and ACW, respectively), parallel contours, and radial contours (concentric circles of lipids), with each showing the corresponding phase contour lines. Each structure is presented for both clockwise (CW) and anti-clockwise (ACW) packing curvatures and the specific structures consistent with those observed in the highlighted domains in (c) are indicated by outlining the structures with dashed boxes and overlaying the structure with the outline of the domains in a position roughly aligned with their observed phases.*

To further investigate the interplay between molecular chirality and mesoscopic structure in these systems, similar experiments with the other DPPC enantiomer (S-DPPC/R-POPC) are performed, with the results shown alongside the R-DPPC data in Fig. 4 (right side). For full comparability, identical preparation procedures and conditions are used for both samples. The obtained in-plane phase image (Fig. 4b) clearly differs for the S-R mixture (cf. R-R), suggesting different molecular structures. From the molecular direction arrows in the highlighted domain (Fig. 4c), it becomes evident that the S-R mixture forms domains with similarly curved packing, but with an opposite curvature direction (ACW). Meanwhile, the corresponding contour line image (also in Fig. 4c) shows that spiralling mesoscopic structures are also formed, but with an inverted spiral direction (thus being part of an ACW (-)-spiral). This inversion of both types of curvature upon enantiomeric exchange confirms that the entire hierarchal structure is dominated by the DPPC chirality. It is important to note that, despite the obvious asymmetry between their domains, the underlying structural motif of both mixtures are perfectly mirror-symmetric. Interestingly, the observed asymmetry only originates from the location of the spiral centre with respect to the domain boundaries. While for the R-enantiomer they fall outside the domains, the S-enantiomer encloses them, making the 'full-spiral' structure clearly visible. On inspection of the phase images in Fig. 4b, these structural properties are clearly common to all domains (see Supplementary Information for further arrow representations).

Based on these findings, it can be concluded that the DPPC-POPC mixtures form condensed DPPC-rich domains circular in shape, but with curved and co-directional molecular spiral arrangements. Furthermore, alongside previous observations on pure DPPC monolayers[7,8,29], this spirally molecular packing seems to be an intrinsic property of DPPC-rich LC domains. We find that both mixtures form domains based on mirror-symmetric spiral structures, but represent different regions within such motifs. This breaks their mirror symmetry. While the exact origin of this discrepancy cannot be assessed in the context of this work, this is clearly a manifestation of enantioselective interactions between the two lipids. The observed differences raise important questions regarding protein binding and domain growth. Specifically, in the R-R mixture, the formed domains are structurally asymmetric, with one side showing concave packing and the other convex. In contrast, the full-spiral structures formed in the S-R mixture are somewhat rotationally symmetric but undoubtedly possess more substantial spatial heterogeneity. It would be interesting to study the impact of this difference on the location and strength of lateral binding to proteins.[44–46] Furthermore, these structures also imply that the domains in the R-R mixture possess varying line tensions around the domain circumference, whereas those in the S-R mixture are expected to be mostly constant. Since the line tension is generally an important factor in growth and aggregation/fusion[47–50], one can expect this to impact the formation, growth, and resulting physicochemical properties of the domains in each enantiomeric mixture.

Experimental evidence for such differences in the formation and growth of the domains in the R-R and S-R mixtures can be found by further analysis of the vSFG images. On inspection of the wider images in Fig. 4a, it is clear that the S-R mixture possesses a lower surface coverage of the LC phase cf. R-R. As detailed in the Supplementary Information, we can quantitatively extract these coverages along with the vSFG signal magnitudes for the LC domains. This allows us to independently derive the relative molecular densities of each lipid within the domains and the degree of orientational order of DPPC. From this analysis, we find that the S-R domains possess ~21% greater densities of DPPC as well as ~22% greater lipid order (i.e., narrower orientational distributions), and exhibit a higher DPPC excess. We can, therefore, conclude that the S-R mixture forms fewer LC domains which have a greater density and increased purity of DPPC, higher molecular order, and a hierarchal structure that includes the spiral centres, possibly suggesting slower formation kinetics. This observation strongly implies that important physicochemical properties such as fluidity, stability, and permeability, clearly differ between these two model membranes. Such substantial enantioselective effects may thus be an important contributor to the evolution of homochiral lipid membranes.

While important questions about the growth mechanism and role of lipid rafts in biology obviously still remain unanswered, the determined hierarchal structure and observed breaking of mirror symmetry in the model systems investigated here represents a crucial insight into the nature of such assemblies and provides a detailed molecular picture of the ripened state within their formation process. Finally, with the presented vSFG microscopy technique, a valuable tool becomes available for the investigation of such systems that can reveal their molecular structure in their full complexity, including in-plane orientations, a critical structural aspect that has previously not been accessible. This advancement in microscopic vibrational imaging hence opens up a promising perspective for further studies surrounding the significance of lipids rafts in our physiology.

# Methods

### Sample Preparation

The lipids 1,2-dipalmitoyl-sn-glycero-3-phosphocholine (R-DPPC, or L-DPPC, >99% purity), 2,3-dipalmitoyl-sn-glycero-1-phosphocholine (S-DPPC, or D-DPPC, >99% purity), and d82-1-palmitoyl-2-oleoyl-glycero-3-phosphocholine (d82-R-POPC, or d82-L-POPC, >99% purity) are obtained from Avanti Polar Lipids (Alabaster, AL, USA) and dissolved in chloroform (99.0-99.4% purity, VWR International GmbH, Darmstadt, Germany) at a concentration of 1 mg ml$^{-1}$. Lipid mixtures are made by combining solutions in a 4:1 volume (mass) ratio of DPPC to POPC.

Mixed lipid monolayers are formed by depositing 20 μL aliquots onto the surface of a PTFE Langmuir-Blodgett trough (MicroTrough G1, Kibron, Helsinki, Finland) dropwise using an Eppendorf pipette. Prior to deposition, the trough is cleaned with both ultrapure water (Milli-Q, 18.2 MΩ·cm, < 3 ppb TOC) and chloroform, filled with ultrapure water, and further cleaned via surface aspiration until the surface pressure was constant within 0.1 mN m$^{-1}$ at full compression. The platinum pressure sensor is flamed and rinsed with both chloroform and ultrapure water to remove any contaminants. Ultra-flat fused silica windows (Korth Kristalle, Altenholz, Germany, 5 mm thick, 25.4 mm diameter, < 2 nm surface roughness) are rinsed with both chloroform and ultrapure water, exposed to UV-ozone (UV/Ozone ProCleaner Plus, BioForce Nanosciences Virginia Beach, VA, USA) for at least 30 mins, and dipped vertically into the sub-phase with a LayerX dipper (Kibron, Helsinki, Finland).

Once clean, with the fused silica substrate dipped, the lipid is deposited and left for a few minutes to let the chloroform evaporate. The film is then compressed to reach a constant pressure of 20 mN m$^{-1}$ with a barrier speed of 10 mm min$^{-1}$ and left further for ~2 hours to equilibrate and to allow the POPC lipid to oxidise due to the ambient ozone.[51] The monolayer is then cast by retracting the substrate through the interface at a speed of 2 mm min$^{-1}$.

### vSFG Microscope

The newly developed heterodyned phase-resolved widefield vSFG microscope used in this work operates fully in the time-domain, utilising the broadband IR and visible output from a Ti:sapphire-based, homebuilt interferometer, the details of which can be found elsewhere.[52] The IR (centred at 3450 nm) and visible (centred at 690 nm) beams are combined with a local oscillator (LO) generated from z-cut quartz in a collinear beam geometry. The three input beams are directed through a custom-drilled hole in a reflective objective (0.78 NA, Pike Technologies, Madison, WI, USA) to reach the sample which is mounted on an automatic XY rotation stage (SR50PP, Newport, Irvine, CA, USA) atop an automatic height controller (TRA25CC, Newport, Irvine, CA, USA). The output beam from the objective is frequency-filtered to isolate the vSFG signals and recorded on an electrically cooled CCD camera (ProEM-HS:1024BX3, Teledyne Princeton Instruments, Trenton, NJ, USA) using paired-pixel balanced imaging (details can be found elsewhere[40]).

### vSFG Image Treatment

vSFG images are obtained by acquiring interferometric images of the sample with 2 fs steps from -300 to 3000 fs time delay between the IR and other input beams (visible and LO).

Subsequent fast Fourier transform (FFT) of the result converts the interferogram images into complex spectral images.

The spectral images presented in this work represent the average of 4-12 individual interferometric scans (12 for the single images shown in Fig. 2, and at least 4 for each rotation for the data presented in Fig. 3 and Fig. 4). Phase and amplitudes of these raw spectral images are normalised by a reference vSFG microscopy measurement of a z-cut quartz crystal at reduced frequency resolution (scanning range from -300 to 300 fs). Each resulting image is then further treated by removing the dark counts as well as a linear baseline (from non-resonant contributions) calculated based on points outside the spectral region of the resonant modes.

*Fitting and Decomposition of vSFG Spectra*

The spatially averaged in-plane and out-of-plane spectra shown in Fig. 3b, as well as the second SVD component shown in Supplementary Information, are deconvoluted into their constituent bands using an in-house Matlab fitting procedure. This fits the spectra with Lorentzian profiles based on the SFG equation[42] using a bounded nonlinear least-squares curve fitting function with known estimates for the resonant frequencies of the relevant $CH_2$ and $CH_3$ bands from the literature.[42] Whilst the quantitative fits arising from such multi-parameter minimisation procedures often yield lower confidences, their purpose in this work is to qualitatively deconvolute the spectra to determine the relative contributions from $CH_2$ and $CH_3$ bands, and thus are more than adequate.

*Singular Value Decomposition (SVD)*

As shown in the manuscript, the rotational analysis yields distinct in-plane and out-of-plane spectra that contain orientationally related contributions. To simplify the obtained 4D dataset and further improve the signal-to-noise ratio, a similar grouping of related spectral contributions can be obtained via principal component analysis. Here, this is performed via a singular value decomposition (SVD) of the entire 4D dataset prior to Fourier transformation of the rotational dimension. As shown in Supplementary Information, the SVD yields only two significant components which highly match the obtained in-plane and out-of-plane spectra presented in Fig. 3. Their in-plane and out-of-plane character is confirmed by comparing their 0-fold and 1-fold magnitude images obtained by Fourier transformation (see Supplementary Information). This analysis step hence both improves the data quality and allows the rotational phase determination to be performed on only a 3D dataset by extracting the in-plane component. The specific molecular direction of the extracted phase is then obtained using the signs of the specific transition dipole vectors in the in-plane SVD component spectrum.

The SVD analysis is performed by truncating the vSFG images along the frequency axis to only include the frequency range of interest, followed by a back-Fourier transform to the time domain to regenerate real (non-complex) data, and undertaking the in-built SVD algorithm in Igor Pro 8 (WaveMetrics, Lake Oswego, OR, USA). The output components are then converted back to the frequency domain via a further Fourier transformation.

*Rotational Analysis*

For the rotational analysis, vSFG images are recorded at sample rotations of at least every 60° across the full range. The treated images are then back-rotated and corresponding pixels identified to overlap the images and create a 4D matrix of surface location (pixel) against spectral frequency and azimuthal (rotational) angle. After SVD, a new 4D matrix of surface location (pixel) against SVD component and azimuthal angle is obtained which is converted into azimuthal frequencies via complex Fourier transform.

## Acknowledgements

We would like to thank T. Khan, the Paarmann research group, M. Krenz along with the technician group, and the FHI mechanical workshop for their contributions towards the microscope design. We also thank A. Paarmann for help in editing the manuscript and the Bluhm research group for use of their LB-trough.

## Author Contributions

A.P.F. and M.T. conceived the project and A.P.F., B.J., and M.T. designed the experiment. Samples were prepared by A.P.F. and B.J., and microscopy data acquired by A.P.F., B.J., and M.T. The subsequent analysis was performed by A.P.F., B.J., and M.T., and all authors discussed the interpretation of the results. A.P.F. drafted the manuscript which was edited by all authors. M.T. and M.W. supervised the work and M.W. acquired funding.

# Supplementary Information

*Singular Value Decomposition*

As mentioned in the methods section, prior to Fourier transformation to convert the 3D vSFG images as a function of azimuthal sample angle to 3D images as a function of azimuthal frequencies, a Principal Component Analysis (PCA) is performed via Singular Value Decomposition (SVD) on the entire 4D dataset. This effectively groups together those spectral components that are linked throughout the entire dataset and leads to a substantial reduction of noise. The results from this SVD analysis are shown in Fig. S1, with the single values of the first 15 SVD components shown in Fig. S1a in decreasing order. From this, it is evident that the majority of the desired information is contained within the first two SVD components, named C1 and C2, as they are the only ones which clearly stand above the slowly decreasing baseline. This is confirmed by noting that no vibrational contrast is observed for the domains beyond the C2 component, as demonstrated in Fig. S2.

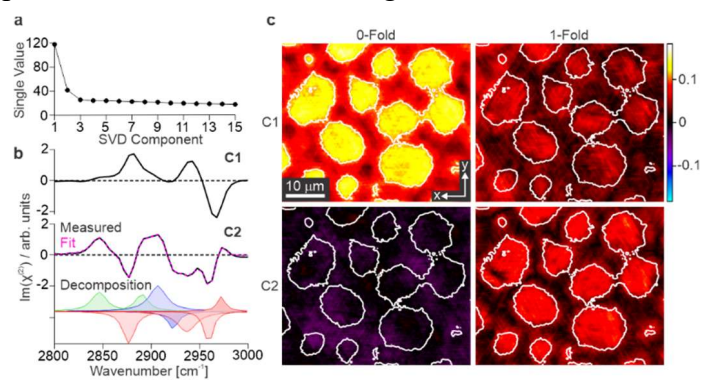

*Fig S1 Singular Value Decomposition (SVD) and rotational analysis. **a**, SVD single values for the first 15 component spectra. **b**, Imaginary parts of the first (C1) and second (C2) SVD component spectra. Included is a deconvolution of the second component into its constituent bands, showing $CH_2$ resonances in green, $CH_3$ resonances in red, and unassigned bands in blue. **c**, Images of the first two SVD components at the 0-fold and 1-fold azimuthal frequencies. As the 0-fold images are entirely real by definition, the real part of the complex FT is shown for each component, whereas the magnitudes are shown for the 1-fold images as these are generally complex. Circular contours highlight the domain locations.*

The spectra of the C1 and C2 components are given in Fig. S1b, only presenting the imaginary (absorptive) parts, where the C2 component is also deconvoluted into its constituent resonances. By comparison to the in-plane and out-of-plane spectra shown in Fig. 3b, it is clear that remarkable similarity exists between these and the SVD component spectra, as expected. Specifically, the C1 component largely indistinguishable from the out-of-plane contribution and the C2 component shows a very similar line-shape and deconvolution profile to the in-plane contribution. This similarity is further confirmed in the rotational analysis of the SVD components given in Fig. S1c which clearly shows the C1 component to be dominated by its 0-fold contribution and similarly the C2 component dominated by its 1-fold contribution.

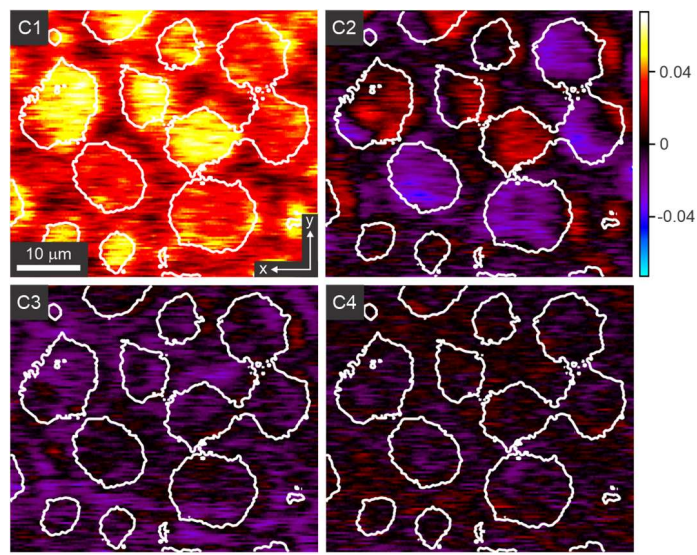

***Fig S2 Singular Value Decomposition (SVD) component images.*** *SVD images of the first four SVD components, C1-C4, for the 0° sample rotation angle. Circular contours highlight the domain locations.*

It is noteworthy, however, that the C1 component does show some (albeit small compared to the 0-fold) intensity at the 1-fold azimuthal frequency, indicating that the SVD has not completely isolated the in-plane and out-of-plane spectra. Instead, they can be generated from a combination of the C1 and C2 components. This is demonstrated in Fig. S3 which overlaps both the in-plane and out-of-plane spectra from Fig. 3b with combinations of the C1 and C2 spectra, showing excellent agreement for both. The out-of-plane spectra is thus generated from a combination of C1 and C2 in roughly a 10:1 ratio, demonstrating clear dominance from the former contribution, and the in-plane spectra from a 5:9 ratio, showing more similar values but still majoritively being described by the C2 component. Nevertheless, as the C2 component is predominantly an in-plane contribution, its rotational phase can be used as a direct measure of the in-plane molecular directionality.

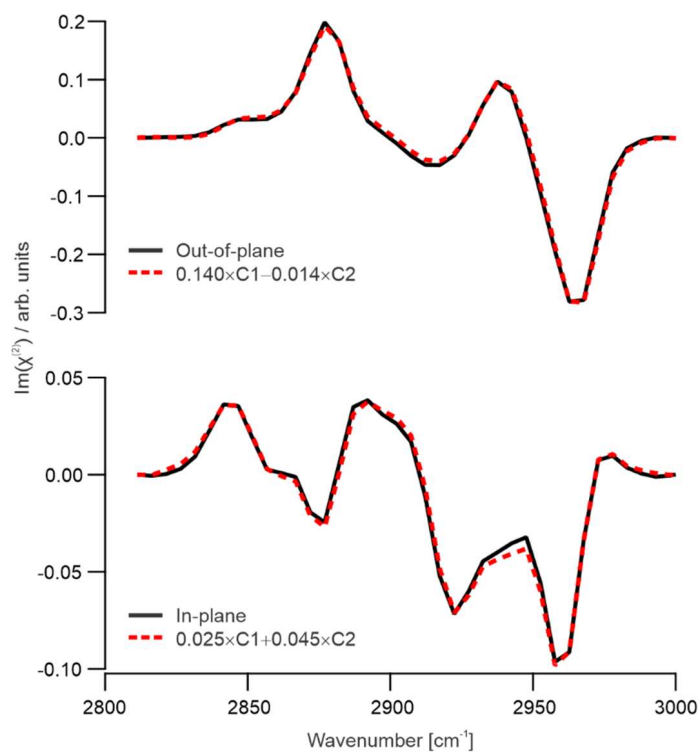

*Fig S3 Comparing SVD component spectra with in-plane and out-of-plane contributions. Spectra obtained from the 0-fold (out-of-plane) and 1-fold (in-plane) azimuthal frequencies (solid black) compared to combinations of the first two SVD component spectra (dashed red).*

### Determining the in-plane molecular direction

As noted above, the C2 SVD component can be used to access the in-plane molecular direction. This is achieved by extracting the rotational phase at the 1-fold azimuthal frequency from the complex Fourier transform of the 3D dataset (pixel vs. azimuthal sample angle) for the C2 component. The direction of this phase is then defined based on the coordinate system in the Laboratory frame, defined with the incident beams be directed within the XZ-plane towards the positive X-direction, which coincides with the coordinate system of the sample frame at 0 deg.

In order to convert this direction of the C2 component into a specific direction based on the molecular structure, one can then utilise the C2 component spectrum, as shown in Fig. S1b. Given that the symmetric $CH_3$ resonance has a very well-defined transition dipole direction and is fairly spectrally isolated, this provides a good marker on which to base the molecular direction. In the C2 spectrum, it can then be observed that the symmetric $CH_3$ stretch has a negative sign. This indicates that the direction of the overall C2 component must be directed opposite to the in-plane contribution of the $CH_3$ symmetric stretch transition dipole. This is the in-plane molecular direction defined both in Fig. 1 and Fig. 4.

### Rotational Analysis Procedure

As discussed in the main text and indicated in Fig. 3, the in-plane molecular orientations are determined from a rotational analysis procedure whereby vSFG images are recorded as a function of azimuthal angle, φ, and subjected to a complex Fourier transform. This converts the azimuthal axis of the 4D dataset to azimuthal (rotational) frequencies, f, as shown schematically in Fig. S4.

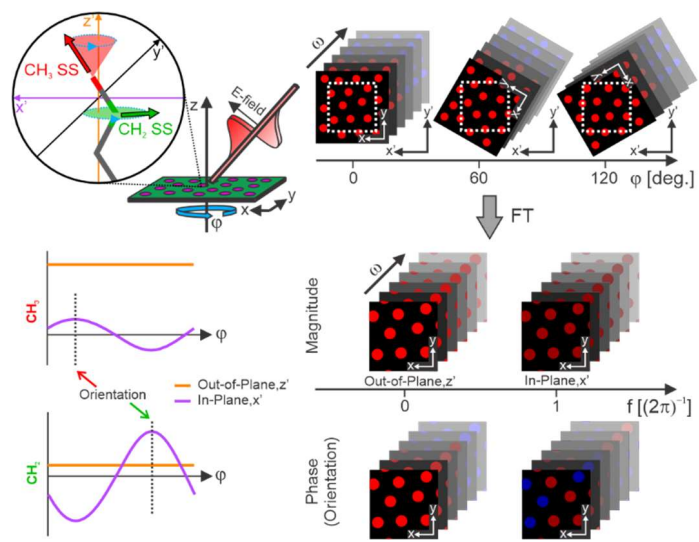

***Fig S4 Rotational analysis procedure.*** *Schematic of the rotational analysis procedure showing the variation in $CH_2$ and $CH_3$ transition dipoles upon rotation and the modulation of their projection along the probed x' and z' directions. Also shown are representations of the four-dimensional dataset that is generated, with SFG microscopy images at different spectral frequencies (ω) recorded at multiple sample rotations (φ), and their conversion into azimuthal frequencies, f, via a Fourier transform (FT), yielding both magnitude and phase data at each azimuthal and spectral frequency.*

By considering the projection of a specific transition dipole vector onto the probed X' and Z' axes, rotation of the sample clearly modulates the in-plane component sinusoidally but leaves the out-of-plane component unchanged. This is also represented in Fig. S4 for the $CH_2$ SS and $CH_3$ SS. This shows that the former (being mostly in-plane) has only a small, constant out-of-plane contribution and a much larger in-plane contribution which oscillates with rotation. By contrast, the $CH_3$ group is mostly directed out-of-plane and thus has a large, constant out-of-plane contribution along with a much smaller, although still non-negligible, contribution from its in-plane projection. Also indicated are phase shifts of the sinusoidal oscillations of the in-plane projections for both modes, demonstrating that the molecular orientation is encoded in the phase of the 1-fold azimuthal frequency.

*Further Arrow Representations*

The in-plane directionality presented in Fig. 4 in the main text shows the full phase map for each enantiomer along with arrow representations for a selected domain to analyse their molecular packing structures. Fig. S5 presents the same phase maps as in Fig. 4b alongside their corresponding arrow representations for each domain within the images. As stated in the main text, it is clear that, whilst there are differences between the domains within each sample, they all show analogous packing structures, disregarding some domains which clearly possess structural defects that likely arise from coalescence or growth abnormalities. Specifically, for the R-enantiomer, each domain possesses a general clockwise turning direction, but only shows modest curvature. By contrast, the domains in the S-enantiomer sample clearly show full spiral structures with anticlockwise turning directions. Specifically, this highlights that the spiralling structure and affinity to form with the spiral centre inside or outside of the domains is common to each well-formed domain and thus represents an intrinsic property of sample associated with their molecular composition.

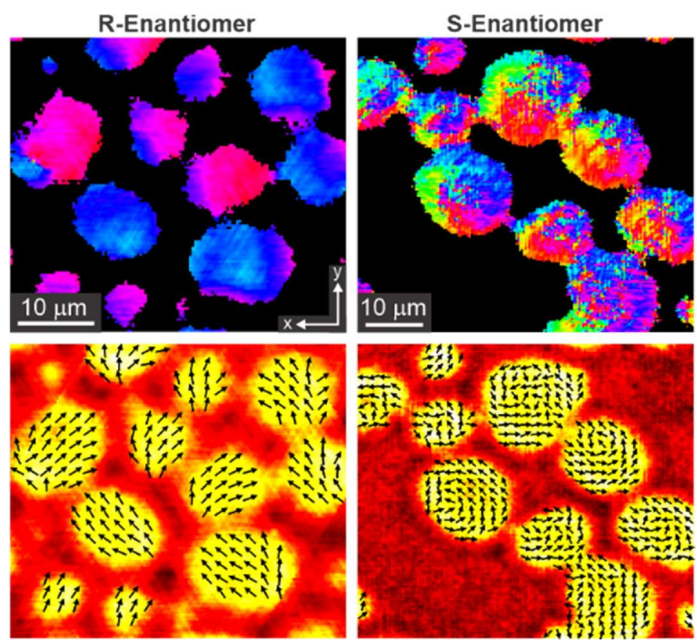

*Fig S5 Phase and arrow representations of domains. Images showing the in-plane phase direction maps as in Fig. 4 with their corresponding molecular direction arrow representations for each domain in the image. Each arrow represents the average phase for a 6×6 pixel region.*

## *Domain Density and Order Calculations*

As discussed in the main text, the R-R and S-R lipid mixtures clearly show significant differences in their in-plane molecular packing. This implies vastly different circumferential line tensions that are likely connected to changes in their formation and growth processes. A further way of assessing these structural differences is from their vSFG responses and relative surface coverages. On inspection of the wider vSFG images presented in Fig. 4a, it becomes evident that the surface coverage of the LC phase in the S-R mixture is significantly lower than for the R-R mixture, despite both containing the same ratio of DPPC to POPC and being formed under equal conditions. This clearly indicates that the formation process of the LC domains must possess some enantioselectivity. As for the comparison of their vSFG responses, these are related to their second-order susceptibilities which represent the macroscopic average of the molecular hyperpolarisabilities. The responses are thus governed by the product of the molecular surface density, $N$, and an order parameter, $O$, that describes the width of the orientational distribution of the molecular transition dipoles.

*Table S1 Parameters used for the Fresnel factor and beam geometry corrections.*

| Parameter | Value |
| --- | --- |
| $n_{air}$ | 1 |
| $n_{FS}$ (3450 nm) | 1.4074 |
| $n_{FS}$ (690 nm) | 1.4555 |
| $n_{FS}$ (575 nm) | 1.4589 |
| $n_{lipid}$ | 1.18 |
| Monolayer Thickness, $h$ / nm | 2 |
| $L_{xx}$ (3450 nm) | 1.0104 |
| $L_{xx}$ (690 nm) | 1.0223 |
| $L_{xx}$ (575 nm) | 1.0232 |
| $L_{zz}$ (3450 nm) | 0.7989 |
| $L_{zz}$ (690 nm) | 0.8083 |
| $L_{zz}$ (575 nm) | 0.8090 |
| Incident Angle, $\theta$ / ° | 36 |

In order to make a meaningful comparison, any orientation-dependence must be removed by extracting the effective magnitude of the response. This can be obtained from the sum-of-squares of the X, Y, and Z contributions that are accessible using the 0-fold and 1-fold azimuthal frequencies. Here, this analysis is performed using the $CH_3$ symmetric stretch which, due to the mostly upright lipid structure, is dominated by its Z-response. It is thus assumed that the 0-fold and 1-fold contributions are dominated by the ZZZ and ZZX (/ZZY) components of the second-order susceptibility, respectively. These responses can thus be corrected for the specific Fresnel factors and incident beam angle, as given in Table S1, to remove any experimental influence and isolate the intrinsic ZZZ and ZZX (/ZZY) terms. The effective magnitudes then become the square-root of the sum-of-squares of these contributions.

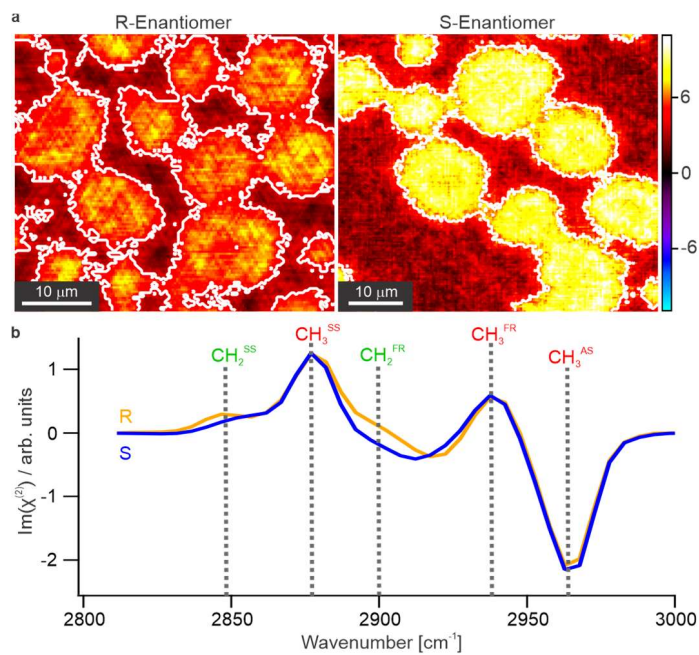

**Fig S6 Order and density of the condensed domains. a**, Images of the symmetric methyl stretch for both enantiomers, corrected for Fresnel factors and beam geometry, represented as the effective magnitude of the transition dipole. **b**, Imaginary part of the C1 component spectra for both enantiomers (R – orange, S – blue) with the main resonances highlighted.

Fig. S6a shows the resulting magnitude maps for both enantiomer samples, where the S-enantiomer clearly yields a stronger response for the LC domains. This shows that this enantiomeric mixture must form domains with an increased DPPC density and/or increased lipid order. To determine the origin of this observed difference, certain parameters can be extracted from the magnitude images in Fig. S6a. Specifically, the total surface area, $A_T$, along with the corresponding area coverages of each phase, $A_{LC}$ and $A_{LE}$, are determined. Thereafter, the vSFG magnitude images in Fig. S6a are integrated over each phase, which must yield the product of the area, number density, and order parameter, i.e., $A_{LC}N_{LC}O_{LC}$ and $A_{LE}N_{LE}O_{LE}$. These parameters are given in Table S2, also normalising the integrated magnitudes based on the areas to yield $N_{LC}O_{LC}$ and $N_{LE}O_{LE}$.

*Table S2 Determined areas and vSFG magnitude parameters from the images in Fig. S6a.*

| Parameter | R | S |
|---|---|---|
| $A_T$ / Px. | 19932 | 33485 |
| $A_{LC}$ / Px. | 11561 | 14399 |
| $A_{LE}$ / Px. | 8371 | 19086 |
| $A_{LC}N_{LC}O_{LC}$ / arb. units | 70432 | 128996 |
| $A_{LE}N_{LE}O_{LE}$ / arb. units | 29384 | 75272 |
| $N_{LC}O_{LC}$ / arb. units | 6.09 | 8.96 |
| $N_{LE}O_{LE}$ / arb. units | 3.51 | 3.94 |

With these parameters in hand, it becomes possible to estimate the ratios of $N_{LC}$ and $O_{LC}$ for the two enantiomers using some simple assumptions. Firstly, it is assumed that the average number density of DPPC across the entire surface (i.e., in both phases) is the same for both enantiomers. This can be summarised mathematically as in Eq. 1. This is a very reasonable assumption as both samples were generated with the same ratio of DPPC to POPC at identical surface pressures, as mentioned previously.

$$\frac{N_{LE}^R A_{LE}^R + N_{LC}^R A_{LC}^R}{A_T^R} = \frac{N_{LE}^S A_{LE}^S + N_{LC}^S A_{LC}^S}{A_T^S} \tag{1}$$

Secondly, it is also assumed that the order parameter in the liquid-expanded (LE) phase is the same for both enantiomers, i.e., as in Eq. 2. Again, this assumption is reasonable since the LE phase yields only small vSFG contributions and is generally considered to be largely structurally disordered and dominated by POPC.

$$O_{LE}^R = O_{LE}^S \tag{2}$$

With these two assumptions, the parameters in Table S2 can be analysed, where the number density ratio in the LE phase can be explicitly determined, as in Eq. 3.

$$\frac{N_{LE}^R}{N_{LE}^S} = \frac{N_{LE}^R O_{LE}^R}{N_{LE}^S O_{LE}^S} = \frac{3.51}{3.94} = 0.89 \tag{3}$$

Similarly, the values for the areas can be implemented in Eq. 1 to yield Eq. 4.

$$0.42 N_{LE}^R + 0.58 N_{LC}^R = 0.57 N_{LE}^S + 0.43 N_{LC}^S \tag{4}$$

Then, utilising Eq. 3, this can be rearranged to yield Eqs. 5 and 6.

$$N_{LC}^S = \frac{\left(0.42 - \frac{0.57}{0.89}\right) N_{LE}^R + 0.58 N_{LC}^R}{0.43} \tag{5}$$

$$\frac{N_{LC}^S}{N_{LC}^R} = 1.35 - 0.51 \frac{N_{LE}^R}{N_{LC}^R} \tag{6}$$

To obtain the desired quantity on the left side of Eq. 6, the number density ratio between the LE and LC phase needs to be determined. Given that the LC phase is known to be DPPC-rich and the LE phase DPPC-poor, clearly this value is substantially less than 1. Nevertheless, as indicated in the main text, there are still spectral contributions from DPPC in the LE phase, showing that this ratio is also non-zero. As a way of achieving a good estimate of the exact value, we can turn to the vSFG responses from both phases. The spectra averaged across each phase presented in Fig. 2c (LC in red, and LE in blue) show a relative amplitude ratio of the CH$_3$ SS of ~0.28. As these spectra are spatially averaged (within their respective regions), they are dominated by the out-of-plane response, and thus combine the molecular density of DPPC with the order parameter in the Z direction. By comparing the vSFG responses from pure DPPC and POPC monolayers given elsewhere in the literature[51], along with their molecular densities, we find that the relative out-of-plane order parameters between the two lipids are almost equal. Therefore, assuming that any DPPC present in the LE POPC-dominated phase has a similar order to POPC, it hence becomes reasonable to take this value of 0.28 as an estimate of the relative DPPC densities.

This then yields the density ratio between the S- and R-enantiomers as in Eq. 7.

$$\frac{N_{LC}^S}{N_{LC}^R} = 1.21 \tag{7}$$

Clearly, therefore, the S-enantiomer sample has LC domains with a greater DPPC density. Using this value, the ratio of order parameters for the LC phase between the two enantiomers can also be determined, as in Eq. 8.

$$\frac{O_{LC}^S}{O_{LC}^R} = \frac{N_{LC}^S O_{LC}^S}{N_{LC}^R O_{LC}^R} \times \frac{N_{LC}^R}{N_{LC}^S} = 1.22 \tag{8}$$

Clearly this suggests that the S-enantiomer domains are both more DPPC-rich (higher density) and more ordered, compared to the R-enantiomer domains.

Using these values for the relative densities of DPPC in each phase (Eqs. 3 and 8), we can then further analyse the data to determine the relative densities of POPC in the LC phase, hereafter defined as $\widehat{N}_{LC}^{R/S}$. For this, we make two further assumptions. Firstly, we assume that the average number density for POPC is the same for both enantiomeric mixtures, as described by Eq. 9.

$$\widehat{N}_T^R = \frac{\widehat{N}_{LE}^R A_{LE}^R + \widehat{N}_{LC}^R A_{LC}^R}{A_T^R} = \frac{\widehat{N}_{LE}^S A_{LE}^S + \widehat{N}_{LC}^S A_{LC}^S}{A_T^S} = \widehat{N}_T^S \tag{9}$$

This follows logically from the same assumption for DPPC expressed in Eq. 1. Secondly, we assume that the total number density of both lipids in the LE phase is also the same for both mixtures, as in Eq. 10.

$$N_{LE}^R + \widehat{N}_{LE}^R = N_{LE}^S + \widehat{N}_{LE}^S \tag{10}$$

This is a reasonable assumption given that similar molecular packing is expected in the LE phase which is known to be dominated by POPC.

The density of POPC in the LC phase can be written in terms of the total, averaged surface density and that in the LE phase as in Eq. 11 for the R-enantiomer.

$$\widehat{N}_{LC}^R = \frac{\widehat{N}_T^R A_T^R - \widehat{N}_{LE}^R A_{LE}^R}{A_{LC}^R} \tag{11}$$

Then, we make use of the known ratio of DPPC to POPC (4:1), such that $\widehat{N}_T^R = \frac{1}{4} N_T^R$ and $\widehat{N}_{LC}^R$ can be written as in Eq. 12.

$$\widehat{N}_{LC}^R = \frac{N_{LE}^R A_{LE}^R + N_{LC}^R A_{LC}^R}{4 A_{LC}^R} - \widehat{N}_{LE}^R \frac{A_{LE}^R}{A_{LC}^R} \tag{12}$$

Similar treatment for the S-enantiomer yields Eqs. 13-15, where the assumptions described by Eqs. 9 and 10 are also exploited.

$$\widehat{N}_{LC}^S = \frac{\widehat{N}_T^S A_T^S - \widehat{N}_{LE}^S A_{LE}^S}{A_{LC}^S} \quad (13)$$

$$= \frac{\widehat{N}_T^R A_T^S - (\widehat{N}_{LE}^R + N_{LE}^R - N_{LE}^S) A_{LE}^S}{A_{LC}^S} \quad (14)$$

$$= \frac{1}{A_{LC}^S} \left[ \frac{N_{LE}^R A_{LE}^R + N_{LC}^R A_{LC}^R}{4 A_T^R} A_T^S - \left(1 - \frac{N_{LE}^S}{N_{LE}^R}\right) N_{LE}^R A_{LE}^S - \widehat{N}_{LE}^R A_{LE}^S \right] \quad (15)$$

Then, taking the difference and using the known values for the areas and DPPC density ratios leads to Eqs. 16-18.

$$\widehat{N}_{LC}^R - \widehat{N}_{LC}^S = \widehat{N}_{LE}^R \left( \frac{A_{LE}^S}{A_{LC}^S} - \frac{A_{LE}^R}{A_{LC}^R} \right)$$

$$+ N_{LE}^R \left( \frac{A_{LE}^R}{4 A_{LC}^R} - \frac{A_{LE}^R}{4 A_{LC}^S} \frac{A_T^S}{A_T^R} + \left(1 - \frac{N_{LE}^S}{N_{LE}^R}\right) \frac{A_{LE}^S}{A_{LC}^S} \right) + \frac{N_{LC}^R}{4} \left(1 - \frac{A_{LC}^R}{A_{LC}^S} \frac{A_T^S}{A_T^R} \right) \quad (16)$$

$$= \widehat{N}_{LE}^R \left( \frac{A_{LE}^S}{A_{LC}^S} - \frac{A_{LE}^R}{A_{LC}^R} \right) + N_{LE}^R \left( \frac{A_{LE}^R}{4 A_{LC}^R} - \frac{A_{LE}^R}{4 A_{LC}^S} \frac{A_T^S}{A_T^R} + \left(1 - \frac{N_{LE}^S}{N_{LE}^R}\right) \frac{A_{LE}^S}{A_{LC}^S} + \frac{N_{LC}^R}{4 N_{LE}^R} \left(1 - \frac{A_{LC}^R}{A_{LC}^S} \frac{A_T^S}{A_T^R} \right) \right) \quad (17)$$

$$= 0.60 \widehat{N}_{LE}^R - 0.54 N_{LE}^R \quad (18)$$

This yields a relation for the difference between the POPC densities in the LC domains of the two enantiomeric mixtures which is dependent on the POPC and DPPC density in the R-enantiomer LE phase (an analogous expression also exists using the S-enantiomer LE phase). Given that the LE phase is dominated by POPC, clearly $\widehat{N}_{LE}^R > N_{LE}^R$. This shows that, while the S-enantiomer domains have greater DPPC density cf. the R-enantiomer i.e., $N_{LC}^S > N_{LC}^R$ (from Eq. 7), they have lower POPC density i.e., $\widehat{N}_{LC}^R > \widehat{N}_{LC}^S$. Evidently, therefore, the LC domains in the S-R mixture have a greater excess (higher purity) of DPPC than those in the R-R mixture.

The other conclusion above where the LC domains in the S-R mixture were shown to possess greater lipid order than those in the R-R mixture follows well with the observation of greater DPPC excess for the S-R domains as DPPC can pack better. This conclusion can also be furthered by comparing the C1 spectra for both enantiomers, which are shown in Fig. S6b. As noted previously, the C1 component largely represents the out-of-plane contribution, thus being dominated by $CH_3$ resonances. To this end, the spectra from both enantiomers are roughly similar. They do, however, have notable differences. Firstly, the S-enantiomer clearly has a reduced $CH_2$ presence, indicated by weaker positive contributions from both the $CH_2$ symmetric stretch and its Fermi resonance. A reduction in out-of-plane $CH_2$ could suggest that the tail structure is more 'upright', possessing a lower tilt angle, such that the $CH_2$ transition dipoles are directed more in-plane. Alternatively, it could also be indicative of a reduction in the density of Gauche defects, indicating a more ordered structure.[42] Secondly, whilst only slight, the S-enantiomer appears to have a lower ratio of the symmetric-to-antisymmetric $CH_3$ stretches, manifesting mostly as a lower intensity of the latter band. Given that the symmetric $CH_3$ transition dipole is directed along the angle of the methyl group, and the antisymmetric modes are perpendicular to this, their out-of-plane contributions can be described by cosine and sine functions of the methyl tilt angle. This renders the symmetric-to-antisymmetric ratio in the C1 component (neglecting the sign difference) a cotangent function, indicating that a reduction in this ratio corresponds to an increase in the methyl tilt angle. Whilst these two observations might appear contradictory, it is important to note that the methyl group is generally considered to be directed away from the molecular tilt direction for even chain-length tails.[25] Therefore, for largely upright chains, an increase in methyl tilt angle aligns with a reduced tail tilt angle, indicating that both observations are consistent with each other. Although these observations alone are insufficient for a conclusive determination of the molecular tilt angle, they provide support to the conclusions from the analysis above which suggests an increased order in the S-enantiomer domains (cf. R-enantiomer).